\documentclass[conference]{IEEEtran}
\IEEEoverridecommandlockouts
\usepackage{cite}
\usepackage{amsmath,amssymb,amsfonts}
\usepackage{algorithmic}
\usepackage{graphicx}
\usepackage{subfig}
\usepackage{textcomp}
\usepackage{xcolor}

\def\BibTeX{{\rm B\kern-.05em{\sc i\kern-.025em b}\kern-.08em
    T\kern-.1667em\lower.7ex\hbox{E}\kern-.125emX}}

\newcommand{\arxivnotice}{%
Accepted for presentation at the 23rd International Bhurban Conference on
Applied Sciences and Technology (IBCAST 2026), Islamabad, Pakistan,
17--20 August 2026. This is the authors' accepted version; the version of
record will appear in the conference proceedings.
\copyright~2026 IEEE. Personal use of this material is permitted. Permission
from IEEE must be obtained for all other uses, in any current or future media,
including reprinting/republishing this material for advertising or promotional
purposes, creating new collective works, for resale or redistribution to servers
or lists, or reuse of any copyrighted component of this work in other works.
}

\begin{document}

\title{Design and Experimental Validation of a Multiband Cross-Polarization Conversion (CPC) Metasurface for Radar Cross Section (RCS) Reduction }

\author{%
\IEEEauthorblockN{Sohaib~Yaqoob~Chaudhry,\; Malik Muhammad Abdullah,\; Salman Liaquat,\; Jamal Haider,\; Umar Khan,\;
Azhar~Hasan}
\IEEEauthorblockA{\textit{National University of Sciences and Technology, Pakistan}\\
syaqoob@cae.nust.edu.pk}
\thanks{\arxivnotice}
}

\maketitle

\begin{abstract}
Radar cross-section (RCS) reduction is a fundamental requirement in modern stealth technology, playing a critical role in the low-observable performance of aerial and naval platforms. Among the various passive RCS reduction strategies, including radar-absorbing materials, absorptive coatings, and artificially engineered surfaces; metasurface-based cross-polarization conversion has emerged as a compelling approach owing to its structural simplicity and low profile. In this work, a single-layer cross-polarization conversion (CPC) metasurface developed on a cost-effective FR4 dielectric substrate ($\varepsilon_{r} = 4.4$, $\tan\delta = 0.02$) is proposed for multiband RCS reduction. The designed structure achieves a polarization conversion ratio (PCR) exceeding $95\%$ at three distinct operating frequencies of $7.8$~GHz, $11.7$~GHz, and $18$~GHz, 
spanning the C-, X-, and Ku-bands, which directly translates into a monostatic RCS reduction exceeding $10$~dBsm at the corresponding bands. The metasurface further demonstrates stable polarization conversion performance under oblique incidence up to $60^{\circ}$, confirming its suitability for wide-angle illumination conditions encountered in practical deployment scenarios. Experimental validation conducted in an anechoic chamber confirms close agreement with full-wave electromagnetic simulations, substantiating the reliability of the fabricated prototype. The proposed design offers a lightweight, low-cost, and high-performance candidate for multiband stealth and low-observable platform applications.
\end{abstract}


\section{Introduction}

Metasurfaces are artificial planar structures that have attracted significant attention in numerous studies due to their unique capability to manipulate the phase, amplitude, and polarization of electromagnetic (EM) waves in the subwavelength regime. In this regard, it is possible to control the behavior of waves by means of changing the geometry of the unit cell, or meta-atom, in a wide spectrum of frequencies ranging from optical to microwave \cite{b1}. Therefore, new opportunities for developing miniature communication and sensing devices emerged.

One of the most studied phenomena associated with metasurfaces is the manipulation of polarization. Isotropic and chiral metasurfaces with intrinsic or extrinsic chirality were used to rotate or transform the polarization state of incident waves \cite{b2, b3, b4, b5}. Reflection-type metasurfaces enabled cross-polarization conversion (CPC) as well as  linear polarization (LP)-to-cross polarization (CP) conversion in various microwave bands \cite{b6, b7, b8}. Split-ring resonators that were initially utilized for polarization manipulation in single-frequency bands were later extended to multiband regimes \cite{b9, b10, b11, b12, b13}.

One of the main motivations for Polarization Conversion Metasurface (PCM) research has been its use for stealth technology by means of Radar Cross Section (RCS) reduction. Some of the recent literature suggests some unique approaches for realizing an ultra wideband (UWB) PCM. For example, Su et al. \cite{paper_su} have designed a 3-bit coding PCM that uses separate control of phase and amplitude for cross polarization reflection to provide diffuse scattering and thus reduce monostatic RCS. Another approach was proposed by Ameri et al. \cite{paper_ameri} where they used double-headed arrows with geometric rotation of unit cells ($90^\circ$, $180^\circ$, and $270^\circ$). This approach allowed them to realize a 10-dB RCS reduction in the fractional bandwidth of 126.5\%. To overcome bandwidth limitations in ultra-thin metasurfaces, Chatterjee et al. \cite{paper_chatterjee} employed air gap units and altered concentric double square rings.

Despite all this progress, achieving stable performance over large angles of incidence has proven to be difficult. Researchers have developed polarization converters that can maintain their properties up to $60^\circ$ in the frequency ranges of C, X, Ku, and Ka bands \cite{b14, b15, b16}. Another approach used has been characteristic mode theory to gain more insight into the nature of unit cell operation, specifically in the X band \cite{b17, b18}. Multifunctional units capable of performing additional tasks in addition to polarization conversion, such as absorbing or transmitting, have also been created \cite{b19, b20}. More recently, Zafar \emph{et al.}~\cite{zafar2025} reported an asymmetric multi-band reflective metasurface that achieves linear and circular polarization conversion across the Ku, K, Ka and U bands with $50^\circ$ angular stability on a $1.6$~mm substrate. 
Nevertheless, achieving simultaneous multi-band cross-polarization conversion in the widely-used C, X, and Ku radar bands, combined with wide-angle stability and a single low-cost substrate suitable for RCS reduction, remains an open challenge.

The current study develops a planar anisotropic metasurface composed of a single layer and asymmetrically structured, that is able to attain a highly efficient CPC effect within three operating frequency bands, namely, C-, X-, and Ku-bands. This particular design offers a polarization conversion ratio (PCR) greater than 95\%, which corresponds to a decrease in monostatic RCS of more than 10~dBsqm. Moreover, this design is capable of achieving stable performance even when the incidence angle is at $60^\circ$, and uses a commercially available and inexpensive FR4 substrate.

\section{Proposed Methodology}

\subsection{Operating Principle of CPC-based RCS Reduction}

RCS reduction using artificial surfaces can be attained through various known methods, such as high absorption coefficient metasurfaces, phase-gradient metasurfaces for scattering energy redistribution into multiple angles, and cross-polarization conversion metasurfaces for polarization modification of backscattered waves. Among them, CPC metasurface method presents a relatively simpler approach involving only one layer structure. In this study, the CPC method will be utilized owing to its simplicity in design and efficient conversion rate to demonstrate RCS reduction in C-, X-, and Ku-bands.

Polarization conversion in a metasurface can be achieved depending on the anisotropy structure designed in each unit cell. When a linearly polarized electromagnetic wave interacts with the metasurface geometry of the anisotropic unit cell gives rise to different and unbalanced responses on the two principal axes ($u$ and $v$ axes). Thus, energy is transferred from the orthogonal components of the reflected wave. To obtain full cross-polarization conversion, CPC metasurface designs should optimize energy transmission such that cross-polarized reflected components $|R_{yx}|$ or $|R_{xy}|$ are maximized up to unity, while co-polarized components $|R_{xx}|$ or $|R_{yy}|$ are minimized down to zero.


The phenomenon of anisotropy in a metasurface unit cell may be explained by the use of plasmonic resonances, whereby each resonance element produces a particular response in the electromagnetic spectrum at a certain frequency. In split-ring resonator (SRR) configurations, the shape of the ring (width of the gap, length of the arm, and interaction with neighboring elements) dictates the behavior of inductance and capacitance (LC), thus determining the resonant frequency where there is maximum polarization conversion \cite{bY}. When several resonators are placed together in a unit cell, there will be mutual interactions resulting in the creation of more resonances, which gives the metasurface ability to work in several discrete frequency bands without adding any layers \cite{bY}. For cross-polarization conversion to take place, the unit cell must be rotated such that the directions of anisotropy are at a $45^\circ$ angle to the polarization direction of the incoming wave. The field is divided equally in both directions, and the phase difference between the $u$ and $v$ directions results in polarization rotation upon reflection \cite{bX}. Once the $180^\circ$ phase difference is reached, the total cross-polarization conversion is achieved \cite{bX}. The function of the ground plane in achieving high efficiency is very vital since it ensures that all the energy transmitted is reflected, ensuring total participation in the conversion process, which yields higher PCR results in the reflection mode metasurfaces than transmission mode metasurfaces \cite{bZ}.

\begin{figure}[t]
    \centering
    \subfloat[]{\includegraphics[width=0.13\textwidth]{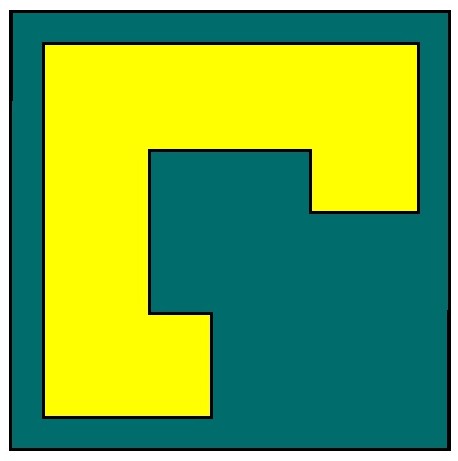}\label{fig:e1}}
    \hfil
    \subfloat[]{\includegraphics[width=0.13\textwidth]{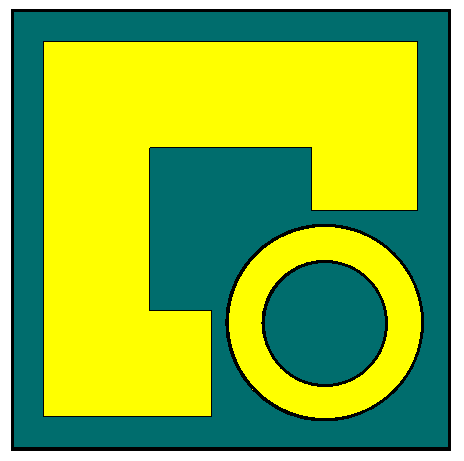}\label{fig:e2}}
    \hfil
    \subfloat[]{\includegraphics[width=0.13\textwidth]{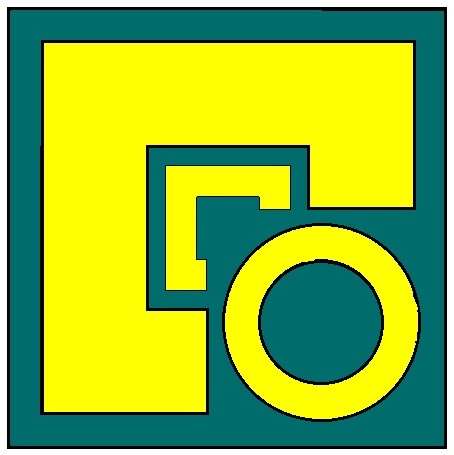}\label{fig:e3}}
    \\
    \subfloat[]{\includegraphics[width=0.19\textwidth]{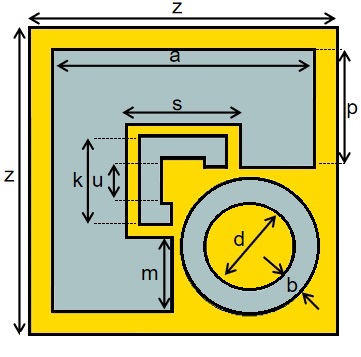}\label{fig:e4}}
    \hfil
    \subfloat[]{\includegraphics[width=0.21\textwidth]{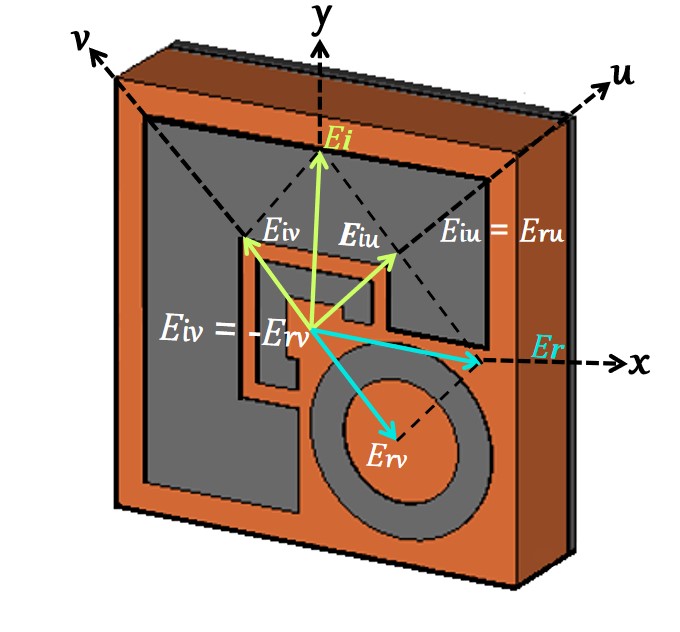}\label{fig:e5}}
    \caption{Unit cell design evolution. (a)~One-side segmented square ring. (b)~Addition of circular ring. (c)~Final structure with extra split ring. (d)~Optimized dimensions. (e)~$u$- and $v$-axis decomposition.}
    \label{fig:evol}
\end{figure}

\subsection{Unit Cell Geometry and Design Evolution}

The geometrical structure of the unit cell was achieved in three progressive design iterations, as shown in Fig.~\ref{fig:evol}, inspired by~\cite{b11}. In the first design iteration, a square ring resonator with a gap on one side of the ring resonator was used to form anisotropy in the structure required for polarization control. Full-wave analysis of this design iteration indicated that the cross-polarized reflection coefficient $R_{xy}$ presents its resonances at 6.41 GHz and 9.73 GHz. The introduction of a circular ring resonator at one corner of the unit cell in the second design iteration resulted in bringing the ratios of magnitudes of co-polarized and cross-polarized reflections close to unity. In the final design iteration, a smaller split-ring resonator with segmentation on only one side of the ring resonator is added to the structure, resulting in the introduction of another CPC resonance in Ku band (15~GHz).The final design of the unit cell, as illustrated in Fig.~\ref{fig:evol} (d), consists of two split ring resonators in addition to the circular ring resonator placed at the corner. The optimal parameters are as follows: $a = 6$~mm, $s = 2.5$~mm, $m = 1.5$~mm, $b = 0.58$~mm, $p = 1$~mm, $z = 7$~mm, $d = 2$~mm. For the design of the unit cell, we have utilized an FR4 substrate material with a dielectric constant $\varepsilon_r = 4.4$, loss tangent $\tan \delta = 0.02$ and thickness of 1.6~mm. In addition to this, the copper metal sheet serves as a ground plane with a conductivity of $\sigma = 5.8 \times 10^7$~S/m to ensure that there will be no transmission of the incoming waves. The overall structure of the metasurface consists of $20 \times 20$ units.

\subsection{Simulation Setup}

Full-wave electromagnetic simulations were performed in the CST Studio Suite using the frequency-domain solver. A single unit cell was modeled with periodic boundary conditions applied along the $x$- and $y$-directions to emulate an infinite array, while Floquet ports were assigned along the $+z$-direction to excite the structure with a normally incident, linearly $x$-polarized plane wave. The bottom face of the substrate was terminated by a perfect electric conductor (PEC) representing the copper ground plane, ensuring zero transmission and confining all incident energy to the reflected mode.

The substrate was modeled as FR4 with relative permittivity 
$\varepsilon_r = 4.4$, loss tangent $\tan\delta = 0.02$, 
and thickness $1.6$~mm. The metallic patterns on the upper 
surface and the ground plane were modeled as copper with 
conductivity $\sigma = 5.8 \times 10^7$~S/m and thickness $0.035$~mm (1~oz). The unit cell periodicity was set to $7$~mm in both lateral directions.

The simulation was carried out for $4$ to $20$~GHz with a frequency resolution of $0.05$~GHz to capture the three resonant features in the C-, X-, and Ku-bands. 
The co-polarized and cross-polarized reflection coefficients, $|R_{xx}|$ and $|R_{yx}|$, were extracted from the Floquet-port S-parameters and were utilized to compute the PCR as well. Angular stability was evaluated by sweeping the incidence angle from $0^\circ$ to $60^\circ$ in increments of $15^\circ$ , with periodic boundary conditions adjusted accordingly to support oblique illumination.

\subsection{Fabrication and Measurement Setup}

\begin{figure}[t]
    \centering
    \subfloat[]{\includegraphics[width=0.20\textwidth]{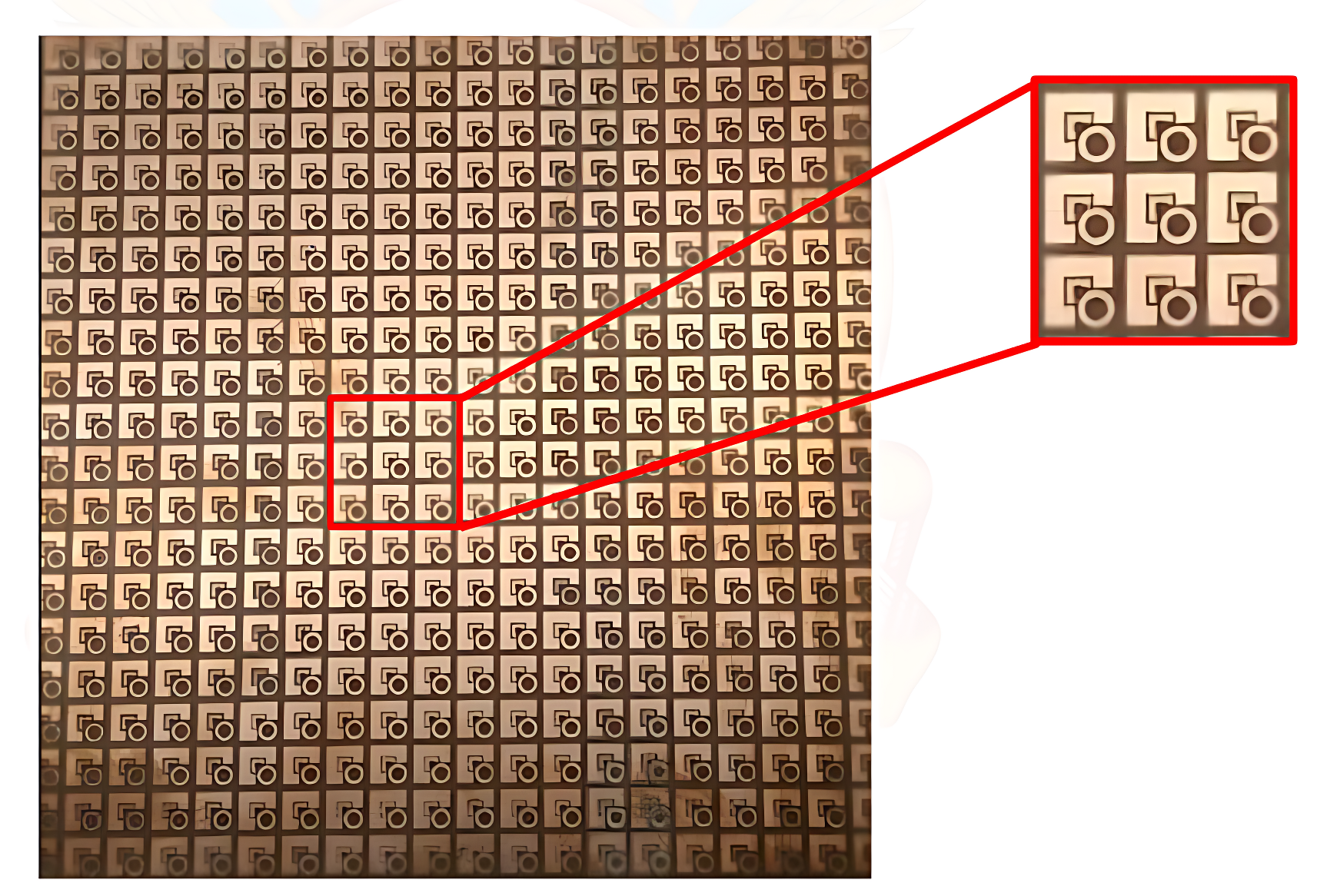}}
    \hfil
    \subfloat[]{\includegraphics[width=0.25\textwidth]{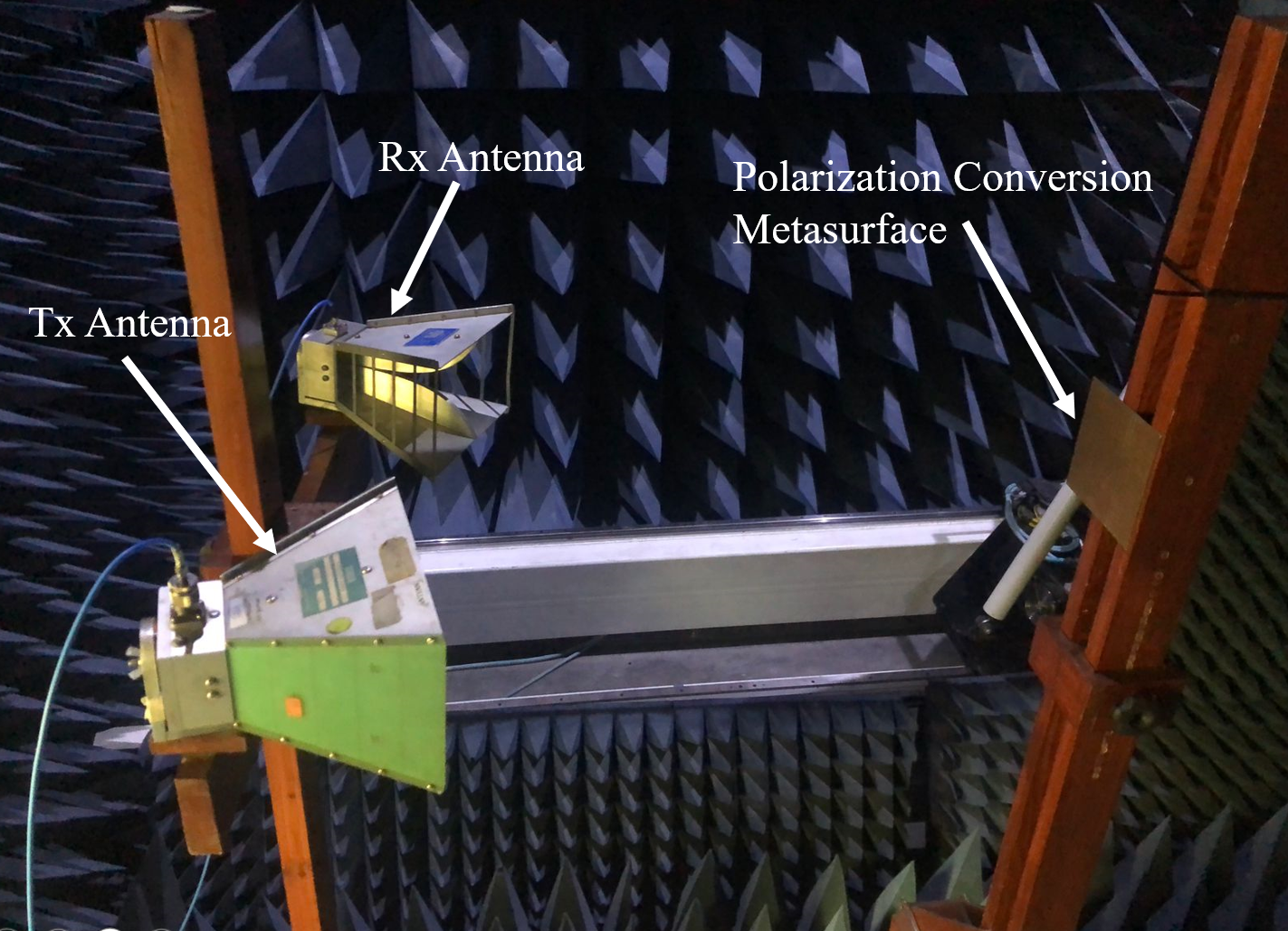}}
    \caption{(a) Fabricated PCM  (b) CPC Measurement setup in the anechoic chamber}
    \label{fig:fab}
\end{figure}

A prototype PCM elements $20 \times 20$ having overall dimensions of $144~\text{mm} \times 144~\text{mm} \times 1.6~\text{mm}$ was 
manufactured on an FR4 substrate using an LPKF ProtoMat M-60 
milling machine. The prototype manufactured is shown in 
Fig.~\ref{fig:fab}(a). Mechanical milling was selected over
chemical etching for rapid prototyping, although it introduces the dimensional error of the order of $\pm 50$~$\mu$m that is considered in the discussion of measured results.

Electromagnetic characterization was performed inside an 
anechoic chamber at NUST, using a vector network analyzer (VNA) to record the $S_{21}$ parameters between two horn antennas. 
A pair of wideband (2-18GHz) standard-gain horn antennas were utilized to span the full frequency range of interest. The transmitting and receiving antennas were positioned at a separation of $0.5$~m from the metasurface in a quasi-monostatic configuration, as illustrated in Fig.~\ref{fig:fab}(b).

To capture both co- and cross-polarized reflection components, the receiving antenna was rotated between the $x$- and $y$-orientations while the transmitting antenna was held in the $x$-polarized state. A reference measurement was first recorded using a flat copper plate of the same dimensions as the metasurface to establish the incident-power baseline; the reflection coefficients of the metasurface were then normalized against this reference. The measured $S_{21}$ magnitude data across all bands was post-processed to extract the cross-polarization conversion (CPC) frequencies and to verify agreement with full-wave simulations.


\section{Results and Discussion}
The performance of the proposed CPC metasurface is evaluated in this section through three complementary analyses:
\begin{itemize}
    \item Simulated reflection and polarization conversion behavior under both normal and oblique incidence
    \item Measured reflection coefficients in the anechoic chamber
    \item Monostatic RCS reduction quantified against a reference metallic plate
\end{itemize}

\subsection{Simulated Reflection Coefficients}

\begin{figure}[t]
\centering
\includegraphics[width=0.5\textwidth]{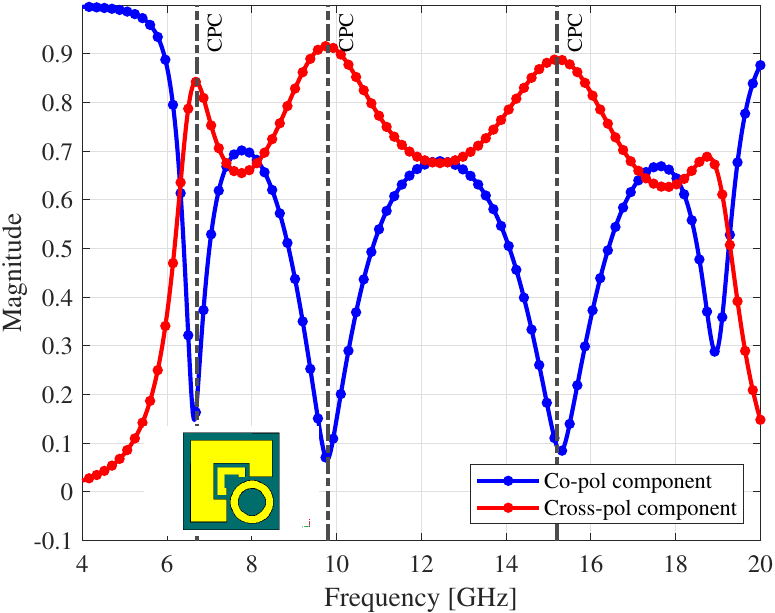}
\caption{Simulated Results of CPC Metasurface in CST}
\label{fig:sim}
\end{figure}

Fig.~\ref{fig:sim} shows the simulated magnitudes of the 
co-polarized $|R_{xx}|$ and cross-polarized $|R_{yx}|$ 
reflection coefficients for a $x$-polarized normally 
incident plane wave. The cross-polarized component exceeds 
$0.8$ at three distinct frequencies, i.e., $6.66$~GHz, $9.75$~GHz, 
and $15.1$~GHz, while the co-polarized component drops to 
near-zero at the same frequencies. This inverse relationship 
between $|R_{xx}|$ and $|R_{yx}|$ confirms the efficient 
redirection of incident energy to the orthogonal 
$y$-polarization across all three operating bands, satisfying 
the necessary condition for cross-polarization conversion.

\subsection{Angular Stability}

\begin{figure}[t]
    \centering
    \subfloat[]{\includegraphics[width=0.44\textwidth]{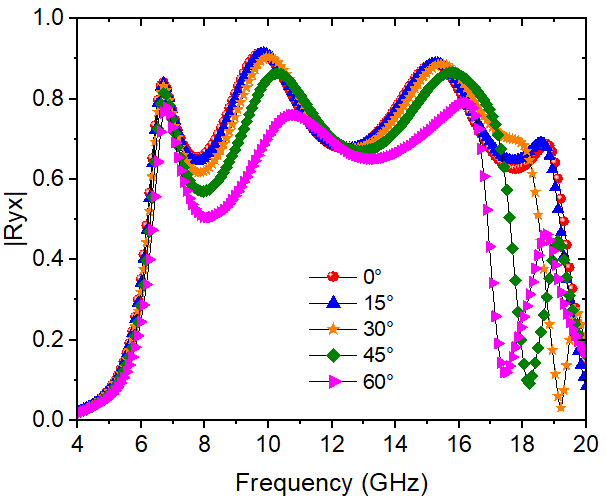}}\\
    \subfloat[]{\includegraphics[width=0.44\textwidth]{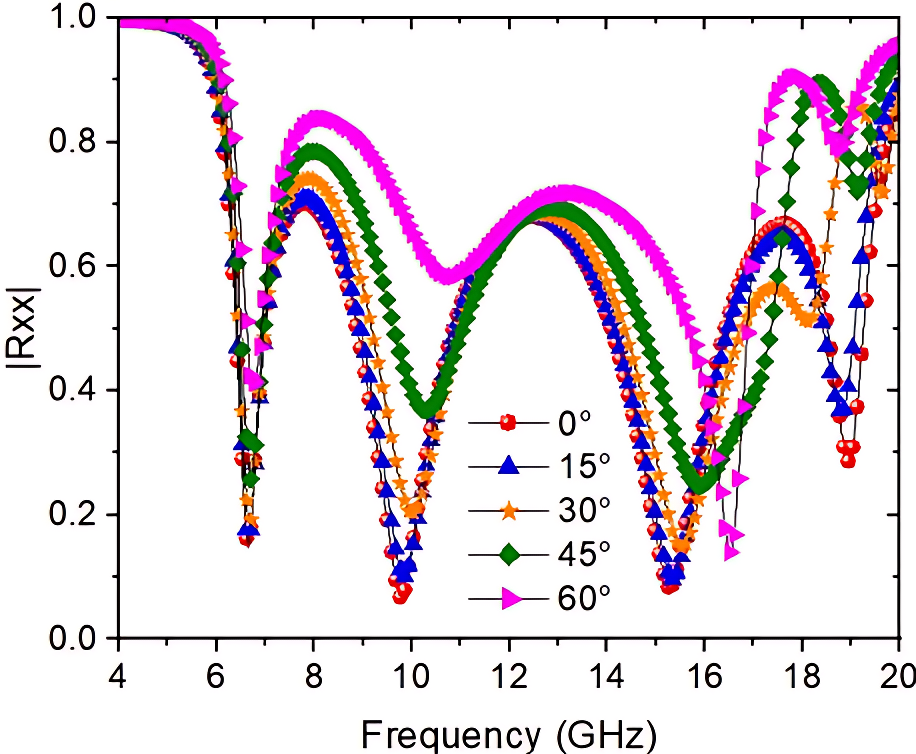}}
    \caption{Angular sweep results (0$^\circ$--60$^\circ$) for (a) cross-pol $|R_{yx}|$ and (b) co-pol $|R_{xx}|$.}
    \label{fig:ang}
\end{figure}

Practical RCS-reduction applications require the metasurface 
to retain its conversion efficiency under oblique illumination, since incident waves rarely arrive at strict normal incidence in deployment scenarios. To assess this, the angle of incidence was swept from $0^\circ$ to $60^\circ$ in $15^\circ$ increments. Fig.~\ref{fig:ang}(a) and (b) show the variation of $|R_{yx}|$ and $|R_{xx}|$, respectively, across this angular range.

The cross-polarized peaks at the three CPC frequencies persist 
with negligible frequency shift and only modest amplitude 
degradation up to $60^\circ$, while the corresponding minima 
of $|R_{xx}|$ remain stable at the same frequencies. This 
indicates that the anisotropic configuration of the unit cell 
preserves polarization conversion under wide-angle incidence, 
making the proposed metasurface suitable for the wide-angle 
illumination conditions encountered in radar and stealth 
applications.

\subsection{Polarization Conversion Ratio}

The polarization conversion ratio (PCR), which quantifies 
the fraction of reflected power residing in the cross-polarized 
component, is defined as
\begin{equation}
\mathrm{PCR} = \frac{|R_{yx}|^2}{|R_{yx}|^2 + |R_{xx}|^2}
\label{eq:pcr}
\end{equation}

where $\mathrm{PCR}$ denotes the polarization conversion ratio, $R_{yx}$ is the cross-polarized reflection coefficient that represents the conversion of an incident x-polarized wave into a reflected y-polarized wave, and $R_{xx}$ is the co-polarized reflection coefficient that represents the reflected component that retains the original x-polarization. The terms $|R_{yx}|^2$ and $|R_{xx}|^2$ correspond to the reflected power in the converted and original polarization states, respectively.

\begin{figure}[t]

\centering

\includegraphics[width=0.46\textwidth]{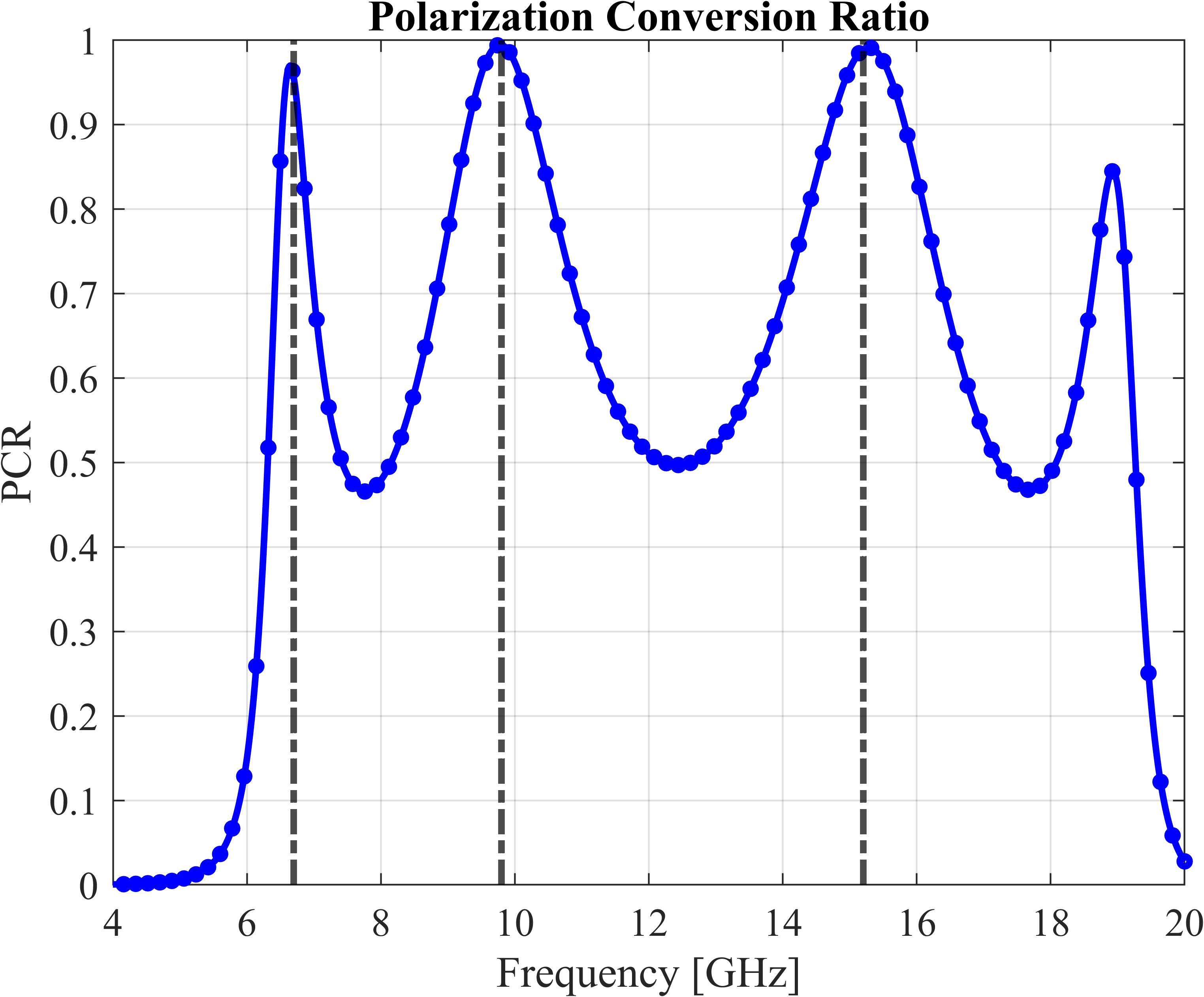}

\caption{Simulated Polarization Conversion Ratio.}

\label{fig:pcr}

\end{figure}

For a CPC-based RCS reduction scheme employing a co-polarized 
monostatic receiver, the achievable RCS reduction is directly 
related to the PCR through
\begin{equation}
\mathrm{RCSR\ (dB)} = -10 \log_{10}(1 - \mathrm{PCR}),
\label{eq:rcsr}
\end{equation}
which predicts a theoretical reduction of approximately 
$13$~dB for $\mathrm{PCR} = 0.95$ and $20$~dB for 
$\mathrm{PCR} = 0.99$. As shown in Fig.~\ref{fig:pcr}, the 
simulated PCR approaches unity at $6.66$~GHz, $9.75$~GHz, 
and $15.1$~GHz, indicating near-complete polarization 
conversion at these three discrete frequencies and 
establishing the theoretical upper bound on the expected 
monostatic RCS reduction. 

\subsection{Measured Reflection Performance}

\begin{figure}[t]
\centering
\includegraphics[width=0.4\textwidth]{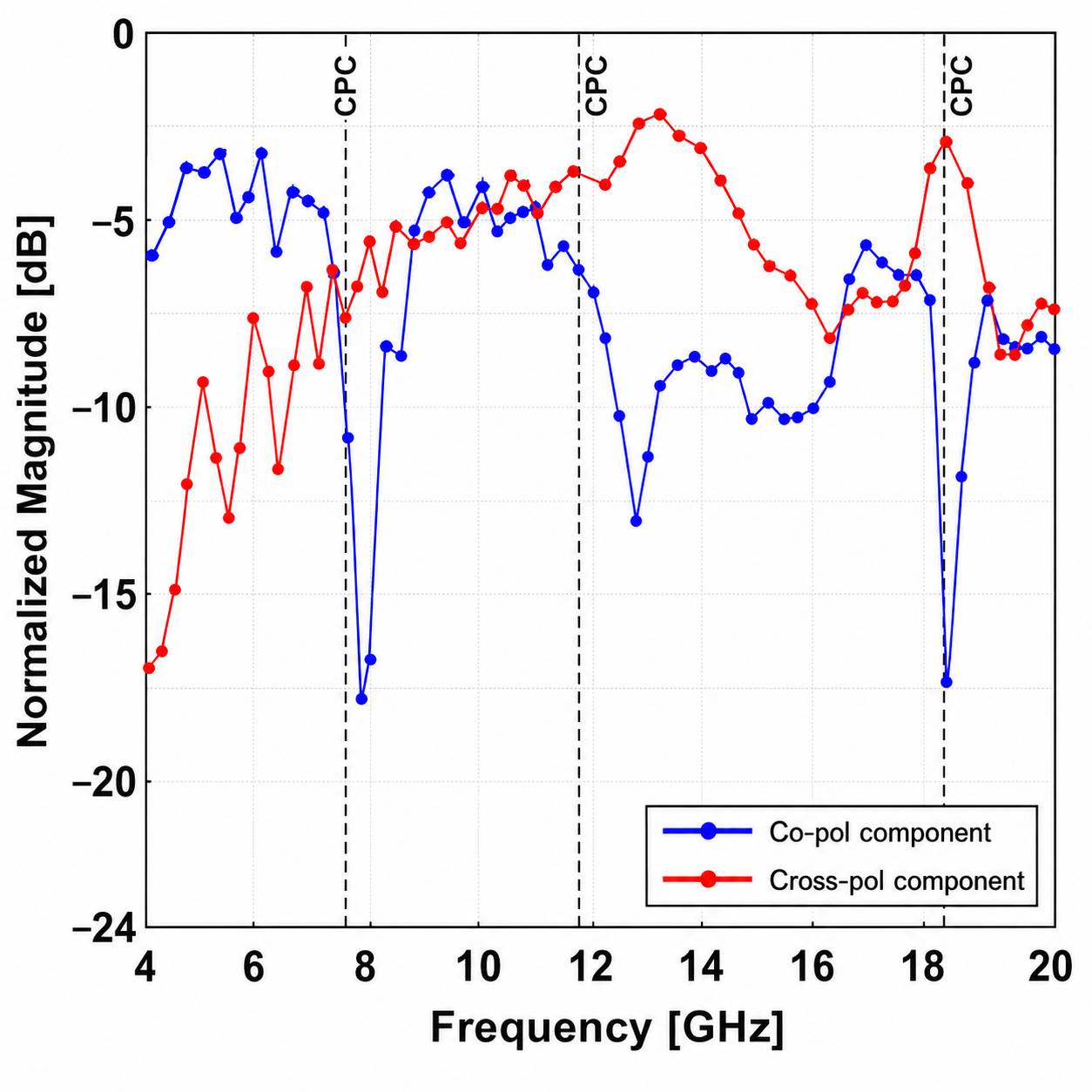}
\caption{CPC Measurement results}
\label{fig:meas}
\end{figure}

The measured magnitudes of the co-polar and cross-polarized 
reflection coefficients across the C-, X-, and Ku-bands are 
shown in Fig.~\ref{fig:meas}. CPC is observed at $7.8$~GHz, $11.7$~GHz, and $18$~GHz, 
where the cross-polarized component reaches its maximum and 
the co-polarized component drops to its minimum. The qualitative 
agreement between simulation and measurement is strong, with 
all three CPC features clearly resolved in the experimental 
data. A frequency offset of approximately $0.5$--$3$~GHz is observed 
between simulated and measured CPC frequencies, with the 
deviation being most prominent in the Ku-band. 


Despite these offsets, the measured response preserves the 
multiband CPC behavior predicted by simulation, confirming the 
effectiveness of the design at all three target bands. Chemical 
etching could be employed in future prototypes to further reduce 
dimensional inaccuracies and improve simulation--measurement 
agreement.

\subsection{Monostatic RCS Reduction}

\begin{figure}[t]
    \centering
    \subfloat[]{\includegraphics[width=0.46\textwidth]{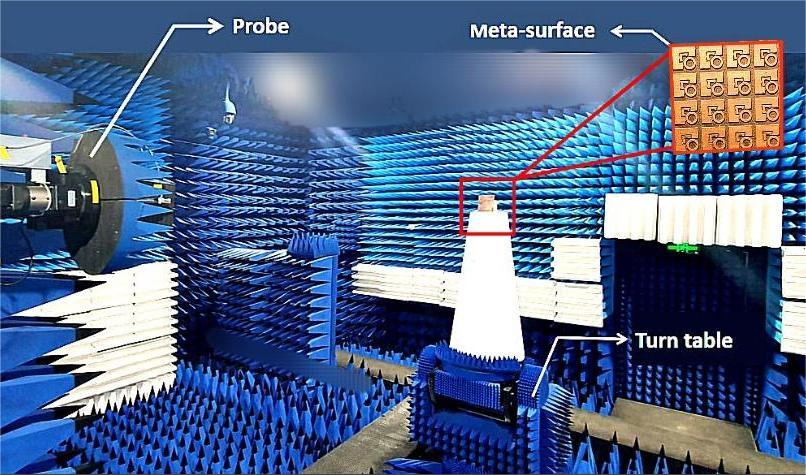}}
    \\[2pt]
    \subfloat[]{\includegraphics[width=0.46\textwidth]{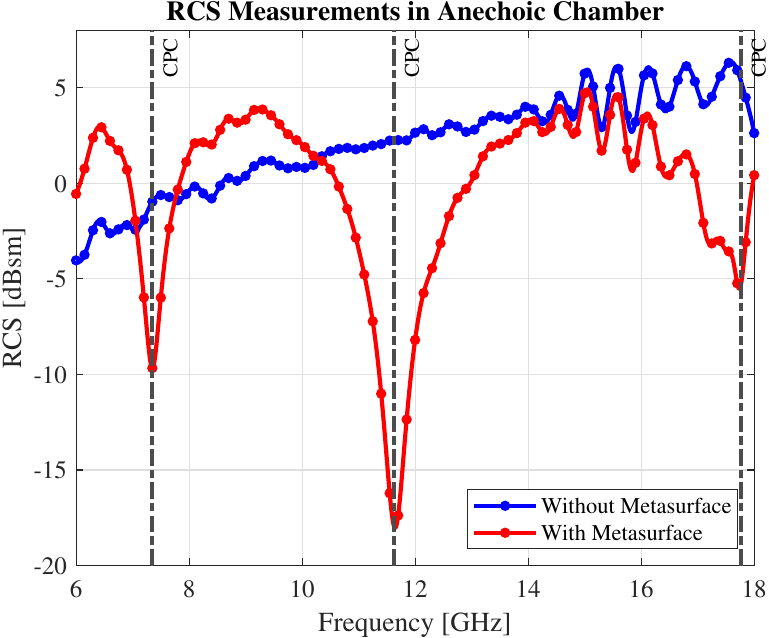}}
    \caption{(a) Anechoic Chamber setup for RCS Measurement and (b) RCS Measurement Results}
    \label{fig:RCS_Meas}
\end{figure}

The RCS-reduction performance was evaluated using a monostatic 
configuration with both transmit and receive antennas in 
vertical--vertical (VV) polarization. The measurement was 
carried out in two stages to establish the baseline performance 
of the designed PCM. The measurement 
arrangement is illustrated in Fig.~\ref{fig:RCS_Meas}(a). First, a reference metallic plate of 
dimensions $10$~cm $\times$ $10$~cm was characterized 
to establish the baseline scattering response, 
shown by the blue curve in Fig.~\ref{fig:RCS_Meas}(b). The 
fabricated metasurface was then placed on the same support 
fixture and characterized under identical conditions, shown by the red curve in Fig.~\ref{fig:RCS_Meas}(b). 

A pronounced reduction in monostatic RCS is observed at 
$7.2$~GHz, $11.5$~GHz, and $18$~GHz, where the metasurface 
achieves cross-polarization conversion efficiencies exceed
$95\%$. The peak measured RCS reduction reaches approximately 
$15$~dB at $11.5$~GHz, $12$~dB at $7.2$~GHz, and $10$~dB at 
$18$~GHz, in close agreement with the values predicted by 
Eq.~\eqref{eq:rcsr} from the simulated PCR. The slight 
underperformance relative to the theoretical upper bound is 
attributed to dielectric losses in the FR4 substrate, 
finite-aperture edge diffraction from the $144$~mm $\times$ 
$144$~mm prototype, and the residual co-polarized component 
caused by imperfect VV polarization alignment of the horn 
antennas. This effect arises from the metasurface's capability 
to alter the polarization of the incident wave, rendering the 
backscattered field largely orthogonal and therefore 
invisible to the co-polarized monostatic receiver. These 
results validate the use of the proposed CPC metasurface as a 
practical RCS-reduction layer at three discrete frequencies 
spanning the C-, X-, and Ku-bands, achieved with a single 
dielectric layer on a low-cost FR4 substrate.

\section{Conclusion}
This work has presented the design, fabrication, and experimental characterization of a compact, single-layer asymmetric anisotropic metasurface for multiband cross-polarization conversion and radar cross-section reduction. The proposed structure achieves CPC with a polarization conversion ratio (PCR) $\ge95\%$ at three distinct operating frequencies of $7.2$~GHz, $11.7$~GHz, and $18$~GHz, collectively spanning the C-, X-, and Ku-bands. This high-efficiency polarization conversion directly translates to a monostatic RCS reduction $\ge$$10$~dB$\cdot$m$^{2}$. The design further maintains stable polarization conversion performance for oblique incidence angles up to $60^{\circ}$, demonstrating robustness under wide-angle illumination conditions. The metasurface was realized on a commercially available and cost-effective FR4 dielectric substrate using a simple dual split-ring resonator geometry. Experimental RCS measurements conducted in an anechoic chamber using a monostatic VV-polarized configuration further validate the simulated predictions, with the measured RCS reduction results showing close agreement with full-wave electromagnetic simulations across all three operating bands.

\end{document}